\documentclass[reprint,pre,twocolumn,superscriptaddress,longbibliography]{revtex4-2}

\usepackage{graphicx}
\usepackage{dcolumn}
\usepackage{bm}
\usepackage{color}
\usepackage[colorlinks=true,allcolors=blue,breaklinks=true]{hyperref}
\usepackage{amsmath} 
\usepackage{physics}

\begin{document}

\title{The structural, vibrational, and mechanical properties of jammed packings of deformable particles in three dimensions}

\author{Dong Wang}
\affiliation{Department of Mechanical Engineering \& Materials Science, Yale University, New Haven, Connecticut 06520, USA}

\author{John D. Treado}
\affiliation{Department of Mechanical Engineering \& Materials Science, Yale University, New Haven, Connecticut 06520, USA}
\affiliation{Integrated Graduate Program in Physical and Engineering Biology, Yale University, New Haven, Connecticut 06520, USA}

\author{Arman Boromand}
\affiliation{Department of Mechanical Engineering \& Materials Science, Yale University, New Haven, Connecticut 06520, USA}

\author{Blake Norwick}
\affiliation{Department of Physics, Yale University, New Haven, Connecticut 06520, USA}

\author{Michael P. Murrell}
\affiliation{Department of Biomedical Engineering, Yale University, New Haven, Connecticut 06520, USA}
\affiliation{Department of Physics, Yale University, New Haven, Connecticut 06520, USA}
\affiliation{Systems Biology Institute, Yale University, West Haven, Connecticut 06516, USA}

\author{Mark D. Shattuck}
\affiliation{Benjamin Levich Institute and Physics Department, The City College of New York, New York, New York 10031, USA}

\author{Corey S. O'Hern}
\email{corey.ohern@yale.edu}
\affiliation{Department of Mechanical Engineering \& Materials Science, Yale University, New Haven, Connecticut 06520, USA}
\affiliation{Integrated Graduate Program in Physical and Engineering Biology, Yale University, New Haven, Connecticut 06520, USA}
\affiliation{Department of Physics, Yale University, New Haven, Connecticut 06520, USA}
\affiliation{Department of Applied Physics, Yale University, New Haven, Connecticut 06520, USA}
 
 \date{\today}

\begin{abstract}
We investigate the structural, vibrational, and mechanical properties of jammed packings of deformable particles with shape degrees of freedom in three dimensions (3D). Each 3D deformable particle is modeled as a surface-triangulated polyhedron, with spherical vertices whose positions are determined by a shape-energy function with terms that constrain the particle surface area, volume, and curvature, and prevent interparticle overlap. We show that jammed packings of deformable particles without bending energy possess low-frequency, quartic vibrational modes, whose number decreases with increasing asphericity and matches the number of missing contacts relative to the isostatic value. In contrast, jammed packings of deformable particles with non-zero bending energy are isostatic in 3D, with no quartic modes. We find that the contributions to the eigenmodes of the dynamical matrix from the shape degrees of freedom are significant over the full range of frequency and shape parameters for particles with zero bending energy. We further show that the ensemble-averaged shear modulus $\langle G \rangle$ scales with pressure $P$ as $\langle G \rangle \sim P^{\beta}$, with $\beta \approx 0.75$ for jammed packings of deformable particles with zero bending energy. In contrast, $\beta \approx 0.5$ for packings of deformable particles with non-zero bending energy, which matches the value for jammed packings of soft, spherical particles with fixed shape. These studies underscore the importance of incorporating particle deformability and shape change when modeling the properties of jammed soft materials.
\end{abstract}

\maketitle

\section{Introduction}


Numerous physical systems are composed of discrete, soft particles that can change shape under applied stress. Examples include collections of emulsion droplets~\cite{paredes13prl, jorjadze11pnas}, colloids~\cite{col}, bubbles~\cite{katgert10epl}, and hydrogel particles~\cite{brodu15natcomm, conleye17sciadv}. These systems display complex, spatio-temporal response to applied deformations, including shear jamming~\cite{cates98prl, bi11nature, bertrand16pre, wang18prl, seto19gm}, shear banding~\cite{unger04prl, moller08pre}, aging~\cite{overlez07pre, shahin11prl, espindola12prl, bonacci20natmat}, and memory formation~\cite{fiocco14prl, royer15pnas, pashineeaax19sciadv}.

Many of the physics-based, theoretical models that are used to investigate the mechanical and vibrational response of soft materials fall into one of two classes: 1) ``soft-particle" models for which the interparticle forces are generated by overlaps between discrete particles of {\it fixed shape}~\cite{durian97pre, ohern03pre, schreck10softmatter} and 2) vertex- or Voronoi-based models~\cite{farhadifar07currbio, bi2015natphys, sussman18prl, merkel19pnas, wang20pnas, merkel18iop} that treat the system as space-filling polygons in two dimensions (2D) or polyhedra in 3D, with interparticle forces determined by shape-energy functions written in terms of the vertices of all polygons in 2D or polyhedra in 3D.

Recently, we introduced the deformable particle (DP) model in 2D that combines optimal features of both classes of models for soft particles~\cite{boromand18prl, boromand19softmatter}. The DP model treats each particle as a discrete object, and thus in contrast to vertex- or Voronoi-based models, the DP model can be used to study systems over a wide range of packing fractions---from isolated particles to confluent systems. In addition, the DP model considers shape-energy functions for each particle individually (through the shape parameter ${\cal A} = p^2/4\pi a$, where $p$ and $a$ are the perimeter and area of the particle, and the bending energy), and thus the DP model provides control over the shape of each deformable particle separately. In previous studies, we investigated the mechanical and vibrational properties of jammed packings of DP particles in 2D with and without bending energy\cite{treado21prm}. We showed that packings of DP particles without bending energy are hypostatic (with fewer contacts than the isostatic value) over the full range of shape parameters, and that the missing contacts are stabilized by low-frequency, {\it quartic} modes of the dynamical matrix. When perturbing the system along quartic modes with amplitude $\delta$, the energy of the system increases as $\delta^4$, rather than $\delta^2$ as for non-quartic modes. Particles with non-zero bending energy undergo a buckling transition when ${\cal A} > {\cal A}^*$, above which the minimal energy shape is not a regular polygon and ${\cal A}^*$ increases with the bending stiffness. Packings of unbuckled particles with ${\cal A} < {\cal A}^*$ are isostatic with no quartic modes. In contrast, packings of buckled particles with ${\cal A} > {\cal A}^*$ possess quartic modes, but we showed that it is difficult to determine how many degrees of freedom are associated with each buckled particle~\cite{santangelo}. The shape degrees of freedom contribute significantly to the vibrational response for packings of DP particles without bending energy, which gives rise to power-law scaling of the ensemble-averaged shear modulus with pressure that differs from that for jammed packings of particles with fixed shape. 

In this article, we develop the DP model in 3D, which considers particles as surface-triangulated polyhedra, and then use it to investigate the structural, mechanical, and vibrational response of jammed packings of deformable particles in 3D. The 3D DP model allows us to determine whether the structural, vibrational, and mechanical properties of jammed packings of deformable particles in 3D are similar to those in 2D, which is important for classifying the critical behavior~\cite{charbonneau15prl:jammingcriticality} of the jamming transition for deformable particle packings. We will show that many of the results for jammed packings of deformable particles are similar in 2D and 3D.  For example, packings of deformable particles with no bending energy possess low-frequency, quartic modes, whose number matches the number of missing contacts from simple contact counting. Also, the pressure-dependent mechanical response varies strongly with the particle deformability; the ensemble-averaged shear modulus scales with pressure as $\langle G\rangle \sim P^{\beta}$ with $\beta \approx 0.75$ for truly deformable particles, whereas $\beta \approx 0.5$ for particles with non-zero bending energy. However, in contrast to the results for 2D, we show that for all non-zero values of the bending energy (i.e. both unbuckled and buckled particles), DP packings in 3D are isostatic at jamming onset and do not possess quartic modes.  

The remainder of the article is organized as follows. In Sec.~\ref{sec:methods}, we describe the shape-energy function for the DP model in 3D and the computational methods used to generate jammed packings of deformable particles in 3D and to calculate the dynamical matrix, density of vibrational modes, stress tensor, and shear modulus for these packings.  In Sec.~\ref{sec:results}, we discuss the results including calculations of the vibrational modes for a single deformable particle (Sec.~\ref{sec:singleparticle}), and the packing fraction and contact number at jamming onset (Sec.~\ref{sec:jamming}), the density of vibrational modes (Sec.~\ref{sec:modes}), the contribution of the shape degrees of freedom to the vibrational modes (Sec.~\ref{sec:shape}), and the mechanical response  (Sec.~\ref{sec:shearmodulus}) of jammed packings of deformable particles in 3D. In Sec.~\ref{sec:conclusions}, we summarize the conclusions and provide promising directions for future research. In addition, we include four Appendices. In Appendix A, we describe the method we employ to decompose the vibrational modes into contributions from the translational, rotational, and shape degrees of freedom of each particle.  In Appendix B, we caclulate the shape parameter distribution for Voronoi tessellations of jammed packings of frictionless spherical particles, as well two types of point processes. In Appendices C and D, we show the influence of adding thermal fluctuations to the compression protocol for generating jammed packings of deformable particles on the properties of jammed packings of deformable particles in two and three dimensions. 

\section{Methods}
\label{sec:methods}

\begin{figure}
    \centering
    \includegraphics[width=\linewidth]{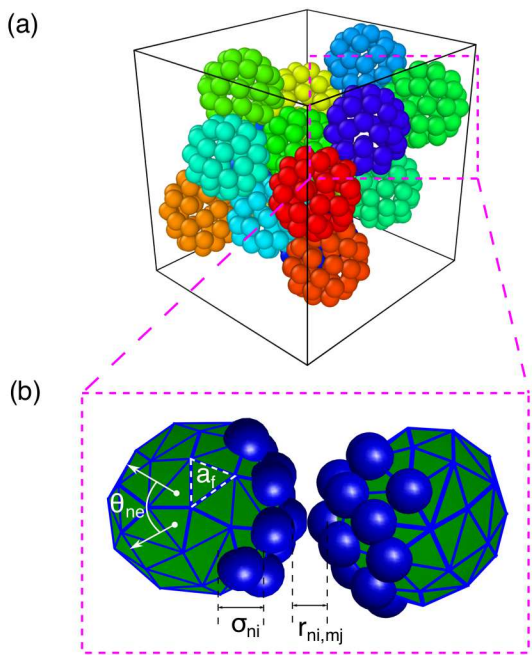}
    \caption{(a) An example jammed packing of DP particles with zero bending energy, $N=16$ particles, $N_v = 42$ vertices, and normalized shape parameter $\widetilde{\mathcal{A}} = 1.04$. (b) Close-up of two particles in (a) to illustrate the definitions of the  surface triangle area $a_{nf}$, the bending angle $\theta_{ne}$, vertex diameter $\sigma_{ni}$, and the inter-vertex separation $r_{ni,mj}$. Several spherical vertices are omitted for clarity.}
    \label{fig:model}
\end{figure}

We model deformable particles in 3D as surface-triangulated polyhedra with $N_v$ vertices as shown in Fig.~\ref{fig:model}. The vertices are connected via Delaunay triangulation, resulting in $N_f$ triangles and $N_e$ edges on the surface of each polyhedron. We characterize the shape of 3D deformable particles using the non-dimensional shape parameter (or \emph{asphericity}) $\mathcal{A} = s^{3/2} / (6 \sqrt{\pi} v)$, where $s$ and $v$ are the total surface area and volume of the particle, respectively. $\mathcal{A}=1$ when the particle is a sphere, and ${\cal A} > 1$ for any non-spherical shape. 

The total potential energy $U$ for a collection of $N$ deformable particles in 3D obeys the following:
\begin{equation}
    \label{eq:shape}
    \begin{split}
    U = & \sum_{n = 1}^{N} \frac{\epsilon_v}{2} \left(\frac{v_n}{v_{0}} - 1\right)^2 + \sum_{n = 1}^{N} \sum_{f = 1}^{N_f} \frac{\epsilon_a}{2} \left(\frac{a_{nf}}{a^f_{0n}} - 1\right)^2 \\
        & + \sum_{n = 1}^{N} \sum_{e = 1}^{N_e} \frac{\epsilon_b}{2} \left( \theta_{ne} \right)^2 + U^{\rm int},
    \end{split}
\end{equation}
where $v_{0}$ is the preferred volume for each particle and $a^f_{0n}$ is the preferred area for the $f$th triangle on the surface of particle $n$. The bending angle $\theta_{ne}$ is the angle between the two unit normals to the triangles that share the $e$th edge on particle $n$. The three coefficients $\epsilon_v$, $\epsilon_a$, and $\epsilon_b$ control fluctuations in the particle volume, surface triangle area, and curvature, respectively. To prevent overlap between deformable particles, we include the purely repulsive, linear spring interaction potential between overlapping spherical vertices on neighboring particles:
\begin{equation}
    \label{eq:inter}
    U^{\rm int} = \sum_{n = 1}^{N} \sum_{m > n}^{N} \sum_{i = 1}^{N_v} \sum_{j = 1}^{N_v} \frac{\epsilon_c}{2} \left(1 - \frac{r_{ni, mj}}{\sigma_{ni, mj}} \right)^2 \Theta \left(1 - \frac{r_{ni, mj}}{\sigma_{ni, mj}}\right),
\end{equation}
where $r_{ni,mj}$ is the distance between the centers of spherical vertices $i$ and $j$ on separate particles $n$ and $m$ and $\sigma_{ni,mj} = (\sigma_{ni} + \sigma_{mj}) / 2$ is the average diameter of the two vertices. The Heaviside step function $\Theta(\cdot)$ enforces that the pairs of vertices only interact when they overlap.

We focus on studies of jammed packings of monodisperse deformable particles in 3D and have verified that they do not possess structural order. To ensure that the particles do not inter-penetrate, we need to have a sufficient number and uniform coverage of the spherical vertices on the surface of each deformable particle. To achieve this, we consider a geodesic polyhedron with $N_v=42$, i.e. the 2nd frequency subdivision of an icosahedron with shape parameter $\mathcal{A}_v = 1.024$.  
For this geodesic polyhedron, there are $N_f=80$ triangular faces: $20$ of the faces have larger area $a^f_{0n} = a_l$, $60$ have smaller area $a^f_{0n}=a_s$, and $a_l/a_s \approx 1.19$.  This geodesic polyhedron also has $N_e=120$ edges, half with larger edge length $l_l$, half with smaller edge length $l_s$, and length ratio $l_l/l_s \approx 1.13$. We choose $\sigma_{ni} = \sigma=l_s$ as the diameter for the spherical vertices.  When providing values of the shape parameter for systems with $\epsilon_b >0$, we provide ${\cal A}$ obtained after minimizing the shape-energy function for an individual particle, not ${\cal A}_0$ defined from $v_0$ and $a_{0n}^f$. (For $\epsilon_b=0$, ${\cal A} = {\cal A}_0$.) Further, we normalize the shape parameter such that ${\widetilde {\cal A}}={\cal A}/{\cal A}_v$.

We consider three important parameter regimes for the DP model in 3D (Eq.~\ref{eq:shape}): 1) Completely deformable particles with $\epsilon_b = 0$; 2) Partially deformable particles with $\epsilon_b > 0$, and 3) ``Rigid" particles for which the relative vertex positions within each particle are fixed (i.e. $\epsilon_v$, $\epsilon_a$, and $\epsilon_b \rightarrow \infty$). For cases (1) and (2), we choose $\epsilon_v \sim \epsilon_a v_{0} \sigma / a_{s}^2 \sim \epsilon_c v_0 / (a_s \sigma) \sim 1$ to achieve comparable area, volume, and vertex-vertex overlap forces near jamming onset. For case (2), we set $\epsilon_b/\epsilon_v = 10^{-4}$ and $10^{-3}$, but the results described below are similar for other values of $\epsilon_b/\epsilon_v$.

To generate jammed packings, we start with a dilute system with packing fraction $\phi = 10^{-3}$, random particle positions in a cubic box with length $L$, and periodic boundary conditions in the $x$-, $y$-, and $z$-directions. We isotropically compress the system by increasing the equilibrium lengths, areas, and volumes of the particles (Eq.~\ref{eq:shape}) in small steps at fixed box length and fixed equilibrium shape parameter.  We start with $\Delta \sigma/\sigma = 10^{-3}$, $\Delta a_s \sim 2 \Delta \sigma/\sigma$, and $\Delta v_0 \sim 3 \Delta \sigma/\sigma$, which corresponds to increases in packing fraction of $\Delta \phi/\phi \approx 10^{-3}$. After each compression step, we use the FIRE algorithm~\cite{fire} to minimize the total potential energy $U$. If the pressure of the energy-minimized packing satisfies $P < P_t$, we compress the system again, followed by energy minimization. If $P > P_t$, we return to the configuration before the most recent compression step and decrease $\Delta \sigma/\sigma$ by a factor of $2$. We continue this process until $1 < P/P_t < 1.1$, where $P_t = 4 \times 10^{-6}$, which yields packings of deformable particles at jamming onset with packing fraction $\phi_J({\widetilde {\cal A}})$.

We calculate the virial stress tensor using
\begin{equation}
    \label{eq:virial}
    \Sigma_{\mu\nu} = \frac{1}{L^3} \sum_{n = 1}^{N} \sum_{m > n}^{N} \sum_{i = 1}^{N_v} \sum_{j = 1}^{N_v} f_{ni,mj,\mu}r_{ni,mj,\nu},
\end{equation}
where $\mu$,$\nu=x$,$y$,$z$, $f_{ni,mj,\mu}$ is the $\mu$th component of the force from vertex $j$ on particle $m$ on vertex $i$ on particle $n$, $r_{ni,mj,\nu}$ is $\nu$th component of the separation vector from vertex $j$ on particle $m$ to vertex $i$ on particle $n$. The pressure is defined as $P = (\Sigma_{xx} + \Sigma_{yy} + \Sigma_{zz}) / 3$. 

To study the vibrational response of jammed packings of deformable particles, we calculate the dynamical matrix:
\begin{equation}
    \label{eq:hessian}
    M_{ni,mj} = \frac{\partial^2 U}{\partial {\vec r}_{ni} \partial {\vec r}_{mj}},
\end{equation}
where ${\vec r}_{ni} = (x_{ni},y_{ni},z_{ni})$ gives the position of the $i$th vertex on particle $n$.
To obtain the elements of the dynamical matrix, we first evaluate $-{\vec f}_{ni} = \partial U / \partial {\vec r}_{ni}$ analytically and then numerically calculate $-\partial {\vec f}_{ni}/\partial {\vec r}_{mj}$ using a finite-difference method on a cubic grid with uniform spacing $10^{-6}L$. We then diagonalize $M_{ni,mj}$ to obtain the ${\cal N} = 3NN_v-3$ non-trivial eigenvalues $\lambda_k$ and corresponding eigenvectors $\vec{V}_k$, with $\vec{V}_k \cdot {\vec V}_{k'} = \delta_{kk'}$ and $k=1,\ldots,{\cal N}$. The eigenfrequencies are given by $\omega_k = \sqrt{\lambda_k/m}$, where all of the vertices have mass $m_i=m$. 

We also measure the packing fraction $\phi_J$ and coordination number $Z_J$ of packings of deformable particles at jamming onset. The packing fraction of a collection of deformable particles is defined as $\phi = \sum_{n=1}^N {\cal V}_{n}/L^3$, where ${\cal V}_n$ is the volume of the $n$th particle. We determine ${\cal V}_n = v_n + \pi N_v \sigma^3/6 - {\cal V}_n^{enc} - {\cal V}_n^{ol}$ by adding the volume $v_n$ of the underlying polyhedron, adding the volume of the spherical vertices, subtracting the volume ${\cal V}_n^{enc}$ of the spherical vertices that is enclosed by the polyhedron, and subtracting the volume ${\cal V}_n^{ol}$ of the overlapping regions between neighboring spherical vertices but outside of the polyhedron.  The volume of the spherical vertices inside the polyhedron is given by ${\cal V}_n^{enc} = \sum_{i=1}^{Nv} \Omega_{ni} \sigma^3 / 24$, where $\Omega_{ni}$ is the solid angle defined by the overlap between the polyhedron and spherical vertex $i$ on particle $n$~\cite{bevis87mathgeo}.  The volume of the overlapped regions between spherical vertices and outside of the polyhedron is given by ${\cal V}_n^{ol} = \sum_{e = 1}^{N_e} \pi (1 - \theta_{ne} / (2\pi)) (2\sigma + l_{ne}) (\sigma - l_{ne})^2 \Theta(1 - l_{ne}/\sigma)/12$, where $l_{ne}$ is the length of the $e$th edge on the $n$th particle.  We also measure the contact number $Z=2N_c/N$ of jammed packings of deformable particles, where $N_c$ is the total number of contacts between distinct pairs of deformable particles.  Note that for two particles $n$ and $m$, multiple vertices on $n$ may overlap multiple vertices on $m$.  However, these are only counted as one contact between particles $n$ and $m$. 

To characterize the mechanical response of jammed packings of deformable particles, we measure the static shear modulus $G$ by applying successive simple shear strains and calculating the resulting shear stress. To generate affine simple shear strain, we shift the $y$-positions of all particle vertices based on their $z$-positions, i.e. the new $y$-positions are given by $y_i' = y_i + \delta \gamma z_i$ with $\delta \gamma = 5 \times 10^{-8}$, we fix their $x$- and $z$-positions, and apply Lees-Edwards boundary conditions. After each shear strain step, we minimize the total potential energy $U$ using FIRE and measure the shear stress $\Sigma = -\Sigma_{yz}$. The shear modulus is given by $G = \partial \Sigma / \partial \gamma$. Finally, note that the length, energy, frequency, and stress scales are provided in units of $L=N^{1/3}$, $\epsilon_c$, $\sqrt{\epsilon_c/m} /L$ and $\epsilon_c /L^3$, respectively.  To assess system-size effects, we study jammed packings with $N=16$, $64$, and $128$.

\section{Results}
\label{sec:results}

In this section, we describe the results from the simulations of jammed packings of deformable particles in 3D. We first study the vibrational response for individual deformable particles with $\epsilon_b = 0$ and $\epsilon_b >0$. As expected, we find that single deformable particles with zero bending energy can change their shape without energy cost, whereas changes in particle shape cost energy for $\epsilon_b>0$. We then investigate the collective structural, vibrational, and mechanical properties in jammed packings of deformable particles. The packing fraction $\phi_J$ and coordination number $Z_J$ at jamming onset increase dramatically with the shape parameter ${\widetilde {\cal A}}$ for completely deformable particles with $\epsilon_b =0$.  However, $\phi_J$ and $Z_J$ do not increase significantly from the values at ${\widetilde {\cal A}}\rightarrow 1$ for packings with nonzero $\epsilon_b$, which is consistent with prior results for packings of frictional, nonspherical particles with rigid shapes. We also show that packings of completely deformable particles possess a large number of low-frequency, quartic eigenmodes of the dynamical matrix, and their number matches the number of missing contacts relative to the isostatic value. In contrast, packings with $\epsilon_b >0$ and rigid-shaped particles are isostatic with no low-frequency, quartic modes. We then decompose the eigenmodes of the dynamical matrix into contributions from the translational, rotational, and shape degrees of freedom of the system. The vibrational response has significant contributions from the shape degrees of freedom over the full range of frequencies for packings of completely deformable particles, whereas there are only large contributions from the shape degrees of freedom at  large frequencies for packings with nonzero $\epsilon_b$. We also show that the ensemble-averaged shear modulus displays power-law scaling with pressure,  $\langle G \rangle \sim P^{\beta}$ for packings of deformable particles, and that the scaling exponent $\beta \sim 0.75$ is larger for packings of completely deformable particles than the value $\beta \sim 0.5$ for packings of particles with non-zero $\epsilon_b$ and rigid, bumpy particles.

\subsection{Single-particle vibrational response}
\label{sec:singleparticle}

\begin{figure}
    \centering
    \includegraphics[width=\linewidth]{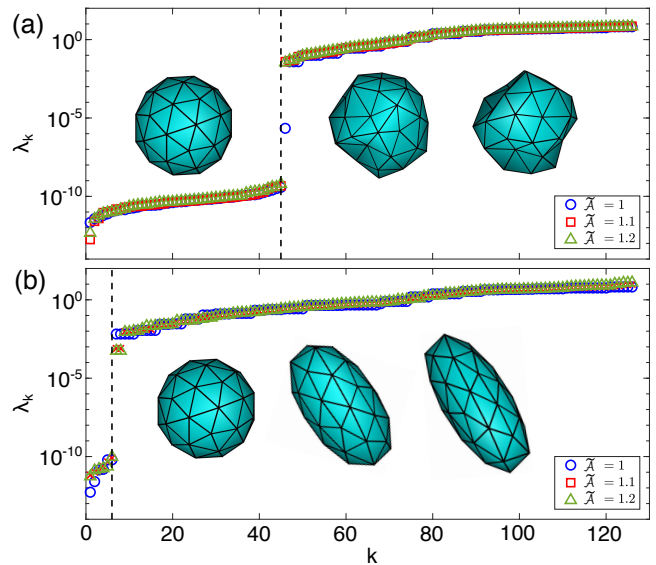}
    \caption{Sorted eigenvalue spectrum $\lambda_k$ (from smallest to largest) for individual deformable particles with (a) $\epsilon_b = 0$ and (b) $\epsilon_b > 0$, and three shape parameters: $\widetilde{\mathcal{A}} = 1$ (blue circles), $1.1$ (red crosses), and $1.2$ (green triangles). The insets show examples for the particle shapes associated with each value of $\epsilon_b$ and $\widetilde{\mathcal{A}}$, with ${\widetilde {\cal A}}$ increasing from left to right. The dashed vertical lines correspond to (a) $k = 45$ and (b) $6$.}
    \label{fig:single}
\end{figure}

For a single deformable particle with $N_v$ vertices, there are $3N_v$ eigenvalues of the dynamical matrix (Eq.~\ref{eq:hessian}). In Fig.~\ref{fig:single} (a), we show the sorted eigenvalue spectrum (from smallest to largest) for a deformable particle with $\epsilon_b =0$ and three shape parameters ${\widetilde {\cal A}}$. For all ${\widetilde {\cal A}}$, we expect $3N_v - N_f -1=45$ zero modes, where $N_f$ gives the number of area constraints for the triangular faces and $-1$ represents the volume constraint.   In Fig.~\ref{fig:single} (a), we show that $\lambda_k \lesssim 10^{-10}$ for $45$ of the eigenvalues, and the remaining $81$ eigenvalues are non-zero with $\lambda_k \gtrsim 10^{-5}$. Deformable particles with $\epsilon_b =0$ can change their shape by moving along eigenvectors associated with these zero eigenvalues. Representative shapes for several ${\widetilde {\cal A}}$ are shown in the inset to Fig.~\ref{fig:single} (a); note that they can possess dimples in their surfaces since $\epsilon_b =0.$

When $\epsilon_b >0$, we add $N_e$ constraints, so that the number of constraints is larger than the number of degrees of freedom. In this case, only rigid translations and rotations of individual particles cost zero energy. As shown in Fig.~\ref{fig:single} (b), deformable particles with $\epsilon_b >0$ possess only $6$ ``zero" eigenvalues $\lambda_k \lesssim 10^{-10}$, corresponding to the three rigid translations and rotations, for all ${\widetilde {\cal A}}$. The remaining eigenvalues are non-zero with $\lambda_k \gtrsim 10^{-4}$.  Thus, deformable particles with $\epsilon_b >0$ can change their shape, but it costs energy. Example minimum energy shapes with $\epsilon_b >0$ are shown in the inset to Fig.~\ref{fig:single} (b).  Note that the shapes at a given ${\widetilde {\cal A}}$ and $\epsilon_b>0$ are more elongated and smooth relative to those at the same  ${\widetilde {\cal A}}$ and $\epsilon_b =0$.

\subsection{Packing fraction and coordination number at jamming onset}
\label{sec:jamming}

\begin{figure}
    \centering
    \includegraphics[width=\linewidth]{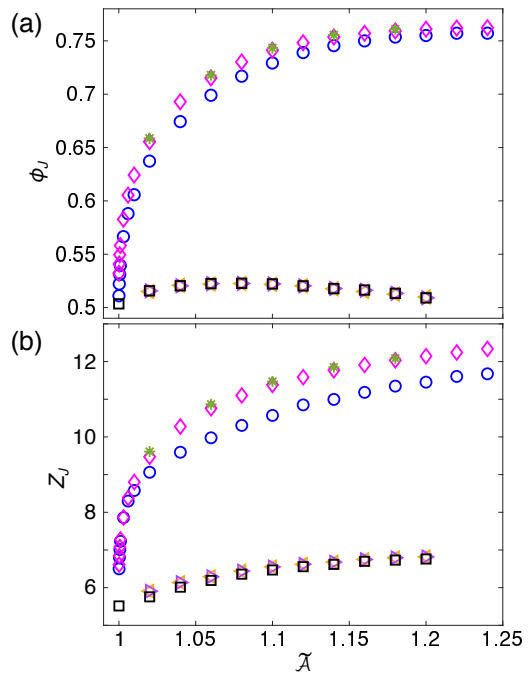}
    \caption{Average (a) packing fraction $\phi_J$ and (b) coordination number $Z_J$ at jamming onset for jammed packings of deformable particles with $N=16$, $\epsilon_b=0$ (circles); $N=64$, $\epsilon_b=0$ (diamonds); $N=128$, $\epsilon_b = 0$ (asterisks); $N=64$, $\epsilon_b = 10^{-3}$ (leftward triangles); $N=64$, $\epsilon_b = 10^{-4}$ (rightward triangles); and $N=64$, rigid shape (squares). The data points are obtained by averaging over $500$ jammed packings.}
    \label{fig:phi}
\end{figure}

In this section, we describe the results for the structural properties (i.e. the packing fraction $\phi_J$ and coordination number $Z_J$) for jammed packings of deformable particles at jamming onset. In Fig.~\ref{fig:phi} (a), we show $\phi_J$ versus ${\widetilde {\cal A}}$ for packings with $\epsilon_b/\epsilon_v =0$, $10^{-4}$, and $10^{-3}$, as well as particles with completely rigid shapes. For completely deformable particles, $\phi_J({\widetilde {\cal A}} \rightarrow 1) \approx 0.50$ and it increases rapidly with ${\widetilde {\cal A}}$, reaching a maximum packing fraction, $\phi_J \approx 0.76$ near, but above ${\widetilde {\cal A}} \gtrsim {\widetilde {\cal A}}^{\dagger} \approx 1.16$.  Note that disordered, jammmed packings of monodisperse, frictional spherical particles have $\phi_J \sim 0.55$ \cite{silbert10sm} in the large-friction limit using the Cundall-Strack model for friction~\cite{cundall}. Thus, the physical roughness of the deformable particles gives rise to more dilute jammed packings in the large-friction limit than those obtained from the Cundall-Strack model.   

The maximum jammed packing fraction is less than $1$ because of the finite size of the spherical vertices. We have shown that the maximum jammed packing fraction increases as the surfaces of the deformable particles become smoother. The shape parameter at which $\phi_J$ reaches its maximum value is similar to the peak value (${\widetilde{\cal A}}^{\dagger} \approx 1.16$) in the probability distribution of shape parameters of the polyhedra generated by Voronoi tessellating jammed, monodisperse frictionless sphere packings as shown in Appendix~B. In Appendices C and D, we show that $\phi_J$ reaches its maximum value at shape parameters closer to ${\widetilde {\cal A}}^{\dagger}$ when the packings are generated by protocols that include thermal fluctuations. 

For any $\epsilon_b >0$, there is a single minimal energy shape at each ${\widetilde {\cal A}}$ and deviations from this shape cost energy.  For this reason, the structural properties (e.g. $\phi_J({\widetilde {\cal A}}))$ for jammed packings of deformable particles with any $\epsilon_b >0$ will differ from those for $\epsilon_b =0$. Further, the structural properties for jammed packings of deformable particles with any nonzero value of $\epsilon_b$ will be similar to those for particles with completely rigid shapes. In particular, in Fig.~\ref{fig:phi} (a), we show that $\phi_J({\widetilde {\cal A}})$ is similar for jammed packings with $\epsilon_b/\epsilon_v = 10^{-4}$ and $10^{-3}$ and with rigid shapes. $\phi_J({\widetilde {\cal A}} \rightarrow 1) \approx 0.50$, $\phi_J({\widetilde {\cal A}})$ increases by a small amount ($\sim 2\%$), reaching a peak near ${\widetilde {\cal A}} \approx 1.08$, and then decreases to $\approx 0.50$ at ${\widetilde {\cal A}} =1.2$. The value at ${\widetilde {\cal A}} \rightarrow 1$ is lower than that found in simulations of frictional, monodisperse spheres using the Cundall-Strack model\cite{silbert10sm} in the infinite-friction limit, but similar to values for random loose packing found in experiments of sequentially deposited rough spheres \cite{menon}.  

The packing fraction at jamming onset for packings of frictionless non-spherical particles typically has a peak near ${\cal A} \sim 1.1$ that is greater than $22\%$ above the value in the ${\widetilde {\cal A}} \rightarrow 1$ limit~\cite{donev07pre,yuan}.  Previous studies of packings of frictional ellipsoids have shown that friction reduces the peak in packing fraction that occurs for small, but finite values of ${\widetilde {\cal A}} -1$ ~\cite{delaney11pre:ellipsoids}.  These prior results are consistent with our observation of a small peak in $\phi_J({\widetilde {\cal A}})$ for deformable particle packings with nonzero $\epsilon_b$.  For both zero and nonzero $\epsilon_b$, $\phi_J({\widetilde {\cal A}})$ does not depend strongly on system size as shown in Fig.~\ref{fig:phi} (a). 

\begin{figure*}
    \centering
    \includegraphics[width=\textwidth]{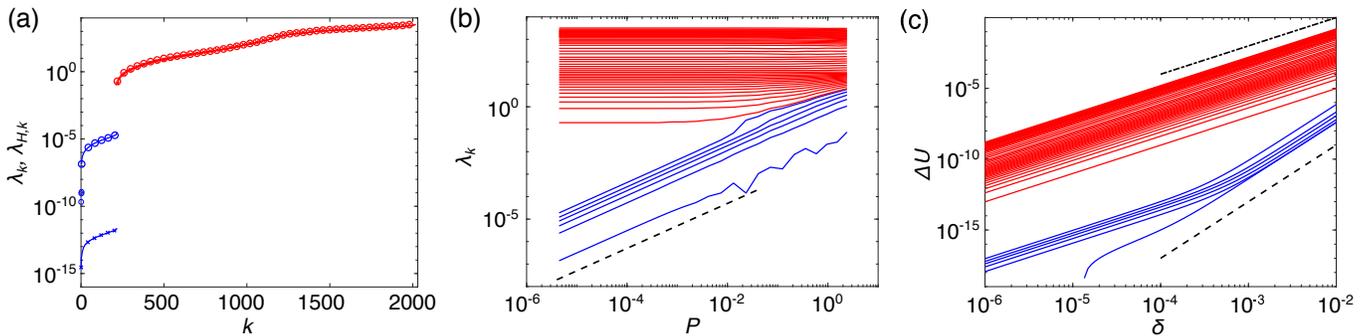}
    \caption{(a) Eigenvalues of the dynamical ($\lambda_k$, circles) and stiffness ($\lambda_{H, k}$, crosses) matrices for a jammed packing of $N=16$ deformable particles with $N_v=42$, $\epsilon_b = 0$, and $\widetilde{\mathcal{A}} = 1.06$, sorted from smallest to largest. This packing has three ``zero"  eigenmodes (with $\lambda_k \lesssim 10^{-9}$, also shown as the first three blue circles), $215$ low-frequency, quartic eigenmodes (with $10^{-7} \lesssim \lambda_i \lesssim 10^{-4}$), and $1798$ quadratic eigenmodes for a total of $3N_v N=2016$ eigenmodes. For quartic modes and quadratic modes, every 30 modes are also shown with circles ($\lambda_k$) and crosses ($\lambda_{H, k}$). (b) The eigenvalues $\lambda_k$ of the dynamical matrix plotted as a function of pressure $P$ during isotropic compression for the same packing in (a). The dashed line has a slope of 1. (c) Change in the total potential energy $\Delta U$ plotted versus the amplitude $\delta$ of the perturbation when the packing in (a) at $P = 4 \times 10^{-6}$ is perturbed along each eigenmode of the dynamical matrix. The dashed (dot-dashed) line has a slope of 4 (2). The blue (red) color of the solid lines in all three panels indicates the quartic (quadratic) modes of the dynamical matrix highlighted by circles in (a).}
    \label{fig:quartic}
\end{figure*}

The coordination number $Z_J =6$ at jamming onset for disordered packings of frictionless spheres\cite{makse00prl,ohern03pre}.  In contrast, $4 < Z_J < 6$ for jammed frictional sphere packings, where the lower value corresponds to the large-friction limit\cite{silbert10sm}. Fig.~\ref{fig:phi} (b) shows results of $Z_J$ for jammed DP packings. We find that $Z_J \approx 5.5$ for ${\widetilde {\cal A}} \rightarrow 1$, which corresponds to the value for packings of frictional spheres with $\mu \approx 0.1$. For completely deformable particles with $\epsilon_b =0$, $Z_J$ increases strongly with ${\widetilde {\cal A}}$, reaching values above $12$ since they can squeeze through the gaps between closely packed particles. $Z_J$ becomes independent of system size for $N \ge 128$. For $\epsilon_b/\epsilon_v =10^{-4}, 10^{-3}$ and completely rigid particles, $Z_J \sim 6$ and it does not increase significantly with ${\widetilde {\cal A}}$.

\subsection{Vibrational response}
\label{sec:modes}

We investigate the vibrational response of jammed packings of deformable particles by calculating the eigenvalues $\lambda_k$ of the dynamical matrix, where $k=1,\ldots,3N_v N$, and the corresponding vibrational frequencies $\omega_k$. We first show the eigenvalue spectrum for jammed packings of completely deformable particles with $\epsilon_b=0$. In Fig.~\ref{fig:quartic} (a), we plot $\lambda_k$ (sorted from smallest to largest) for $N = 16$, $N_v=42$, and $\widetilde{\mathcal{A}} = 1.06$. Apart from the three ``zero" eigenvalues (with $\lambda_k \lesssim 10^{-9}$) from the periodic boundary conditions, we find two distinct bands in the eigenvalue spectrum: one with $215$ eigenvalues that satisfy $10^{-7} \lesssim \lambda_k \lesssim 10^{-4}$ and the other with $1798$ eigenvalues that satisfy $10^{-1} \lesssim \lambda_k \lesssim 10^3$. To better understand the low-frequency band, we investigate the pressure dependence of $\lambda_k$ as the jammed packings are isotropically compressed above jamming onset.   The higher-frequency eigenvalues are nearly independent of pressure $P$, whereas the low-frequency eigenvalues increase linearly with $P$, as shown in Fig.~\ref{fig:quartic} (b). Thus, these low-frequency eigenvalues of the dynamical matrix tend to zero in the $P \rightarrow 0$ limit.

The observation of pressure-dependent eigenvalues of the dynamical matrix for packings of completely deformable particles raises the question of whether these packings are mechanically stable in the zero-pressure limit. To address this question, we perturb the packings by an amplitude $\delta$ in the direction of each eigenmode $\vec{V}_k$:
\begin{equation}
    \label{eq:perturb}
    {\vec R} = \vec{R}_{0} + \delta \vec{V}_k,
\end{equation}
where ${\vec R}$ represents the positions of all vertices on all particles in the perturbed packing and ${\vec R}_0$ represents those in the original packing. In Fig.~\ref{fig:quartic} (c), we show that the change in the total potential energy $\Delta U=U({\vec R})-U({\vec R}_0)$ increases quadratically with $\delta$ for perturbations along eigenmodes in the higher-frequency band. However, for perturbations along the low-frequency eigenmodes, $\Delta U \sim \delta^2$ for small $\delta$ and $\Delta U \sim \delta^4$ for large $\delta$. Based on the results in Fig.~\ref{fig:quartic} (b), the crossover, $\delta^*$, that separates the $\delta^4$ and $\delta^2$ scaling regimes decreases as $\sqrt{P}$. Thus, in the $P \rightarrow 0$ limit, the potential energy increases quartically, not quadratically, with the perturbation amplitude in these directions. These ``quartic" eigenmodes of the dynamical matrix have also been observed in jammed packings of rigid non-spherical particles \cite{mailman09prl, schreck10softmatter, schreck12pre, vanderwerf18pre}.

We further investigate the existence of quartic eigenmodes of the dynamical matrix for packings of completely deformable particles by decomposing the dynamical matrix into contributions from the stiffness and stress matrices, $M = H - S$~\cite{donev07pre, schreck12pre, treado21prm}. The total potential energy for completely deformable particles has three terms, $U = U^v+U^a+U^{\rm int}$ defined in Eqs.~\ref{eq:shape} and~\ref{eq:inter}, and thus the stiffness and stress matrices have three terms, $H=H^v+H^a+H^{\rm int}$ and $S = S^v+S^a+S^{\rm int}$.
The stiffness matrices for each of the three terms are given by: 
\begin{equation}
    \label{eq:stiffnessmatrix}
    H^v_{ni, mj} = 
    \begin{cases}
        \frac{\partial^2 U^v}{\partial v_n^2} \frac{\partial v_n}{\partial {\vec r}_{ni}} \frac{\partial v_n}{\partial {\vec r}_{nj}},& \text{if } n = m\\
        0,              & \text{otherwise}
    \end{cases},
\end{equation}
\begin{equation}
    \label{eq:stiffnessmatrix_a}
    H^a_{ni, mj} = 
    \begin{cases}
       \sum\limits_{f=1}^{N_f} \frac{\partial^2 U^a}{\partial a_{nf}^2} \frac{\partial a_{nf}}{\partial {\vec r}_{ni}} \frac{\partial a_{nf}}{\partial {\vec r}_{nj}},& \text{if } n = m\\
        0,              & \text{otherwise}
    \end{cases},
\end{equation}
and
\begin{equation}
    \label{eq:stiffnessmatrix_c}
    H^{\rm int}_{ni, mj} =  \frac{\partial^2 U^{\rm int}}{\partial r_{ni,mj}^2} \frac{\partial r_{ni,mj}}{\partial {\vec r}_{ni}} \frac{\partial r_{ni,mj}}{\partial {\vec r}_{mj}}.
\end{equation}
The stress matrices for each of the three terms are given by: 
\begin{equation}
    \label{eq:stressmatrix}
    S^v_{ni, mj} = 
    \begin{cases}
        -\frac{\partial U^v}{\partial v_n} \frac{\partial^2 v_n}{\partial {\vec r}_{ni} \partial {\vec r}_{nj}},& \text{if } n = m\\
        0,              & \text{otherwise}
    \end{cases},
\end{equation}
\begin{equation}
    \label{eq:stressmatrix_a}
    S^a_{ni, mj} = 
    \begin{cases}
        -\sum\limits_{f=1}^{N_f} \frac{\partial U^a}{\partial a_{nf}} \frac{\partial^2 a_{nf}}{\partial {\vec r}_{ni} \partial {\vec r}_{nj}},& \text{if } n = m\\
        0,              & \text{otherwise}
    \end{cases},
\end{equation}
and
\begin{equation}
    \label{eq:stressmatrix_c}
    S^{\rm int}_{ni, mj} = -\frac{\partial U^{\rm int}}{\partial r_{ni,mj}} \frac{\partial^2 r_{ni,mj}}{\partial {\vec r}_{ni} \partial {\vec r}_{mj}}.
\end{equation}
 
\begin{figure}
    \centering
    \includegraphics[width=\linewidth]{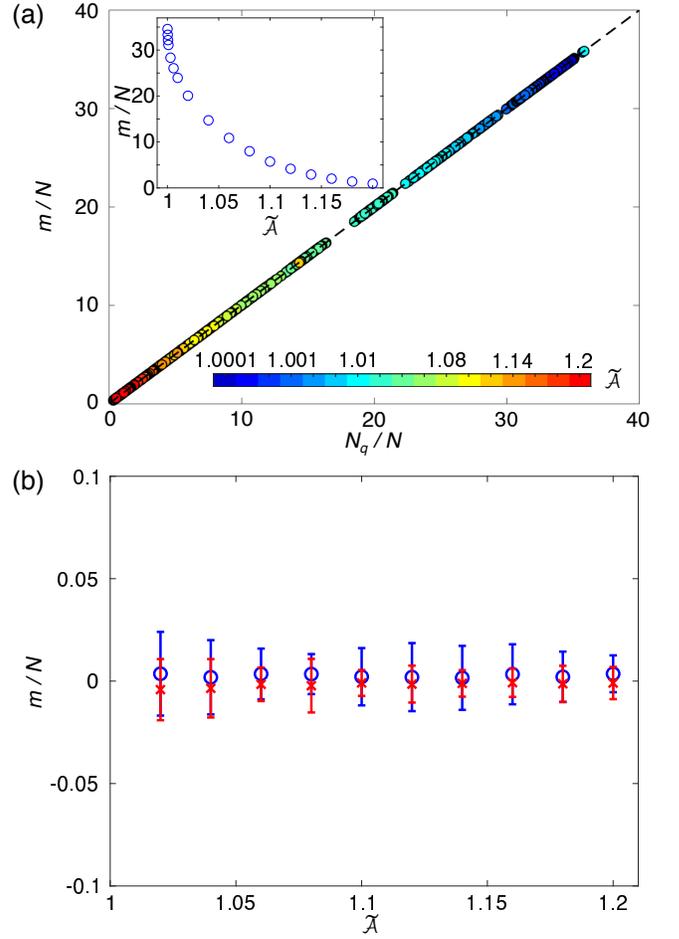}
\caption{(a) Average number of missing contacts per particle $m/N$ (relative to the isostatic value) plotted versus the number of quartic modes per particle $N_q/N$ for $N = 64$ packings of deformable particles with $\epsilon_b = 0$.  The colors of the symbols indicate the value of the shape parameter from ${\widetilde {\cal A}} =1$ (cyan) to $1.2$ (magenta). The dashed line indicates $m/N= N_q/N$. The inset shows $m/N$ versus ${\widetilde {\cal A}}$ for the same data in the main plot. (b) The number of missing contacts $m/N$ plotted versus $\widetilde{\mathcal{A}}$ for $N = 64$ jammed packings of deformable particles with $\epsilon_b/\epsilon_v = 10^{-3}$ (blue circles) and rigid bumpy particles (red crosses) with same values of $N_v$ and $\widetilde{\mathcal{A}}$. In both panels, the data were obtained by averaging over $500$ packings.}
    \label{fig:isostaticity}
\end{figure}

The number of non-zero eigenvalues $\lambda_{H,k}$ for the stiffness matrix $H$ provides the number of degrees of freedom that are linearly constrained, while the number of non-zero eigenvalues $\lambda_k$ for the dynamical matrix $M$ provides the total number of constrained degrees of freedom. For jammed packings of completely deformable particles, we find that the number of ``zero" eigenvalues of the stiffness matrix (with $\lambda_{H,k} \lesssim 10^{-12}$) matches the number of quartic eigenvalues of the dynamical matrix plus the three trivial zero modes for periodic boundary conditions, as shown in Fig.~\ref{fig:quartic} (a). (Calculating the zero eigenvalues of the stiffness matrix provides a straightforward method for independently identifying the quartic eigenmodes of the dynamical matrix.) We find that the number of missing contacts relative to the isostatic value, $m = N^{\rm iso}_c -N_c=N_q$ with $N_c^{\rm iso}=3N_v N-2$, matches the number of quartic modes $N_q$. This relationship holds for jammed packings of completely deformable particles over the full range in $\widetilde{\mathcal{A}}$ studied, as shown in Fig.~\ref{fig:isostaticity} (a). From Fig.~\ref{fig:phi} (b), we know that $N_c$ increases with ${\widetilde {\cal A}}$, and thus the number of missing contacts decreases with ${\widetilde {\cal A}}$, reaching zero for ${\widetilde {\cal A}} \gtrsim 1.16$ as shown in the inset to Fig.~\ref{fig:isostaticity} (a).

Jammed packings of deformable particles with nonzero $\epsilon_b$ possess only a single band of quadratic eigenmodes, and are isostatic with $m \approx 0$ for all shape parameters studied, as shown in Fig.~\ref{fig:isostaticity} (b) for the specific case of $N=16$ packings with $\epsilon_b/\epsilon_v = 10^{-3}$.  Similar results are found for packings of rigid bumpy particles with the same $N_v$ and ${\widetilde {\cal A}}$. The fact that 3D jammed packings of rigid bumpy particles are isostatic is consistent with prior studies of jammed packings of rigid bumpy particles in 2D\cite{papanikolaou13prl}. In contrast, we showed previously that jammed packings of ``buckled" deformable  particles with $\epsilon_b > 0$ in 2D are {\it hypostatic} with $m=N_q$ quartic eigenmodes of the dynamical matrix\cite{treado21prm}. These results emphasize an important distinction between jammed packings of deformable particles in 2D versus 3D. 

\begin{figure*}
    \centering
    \includegraphics[width=\textwidth]{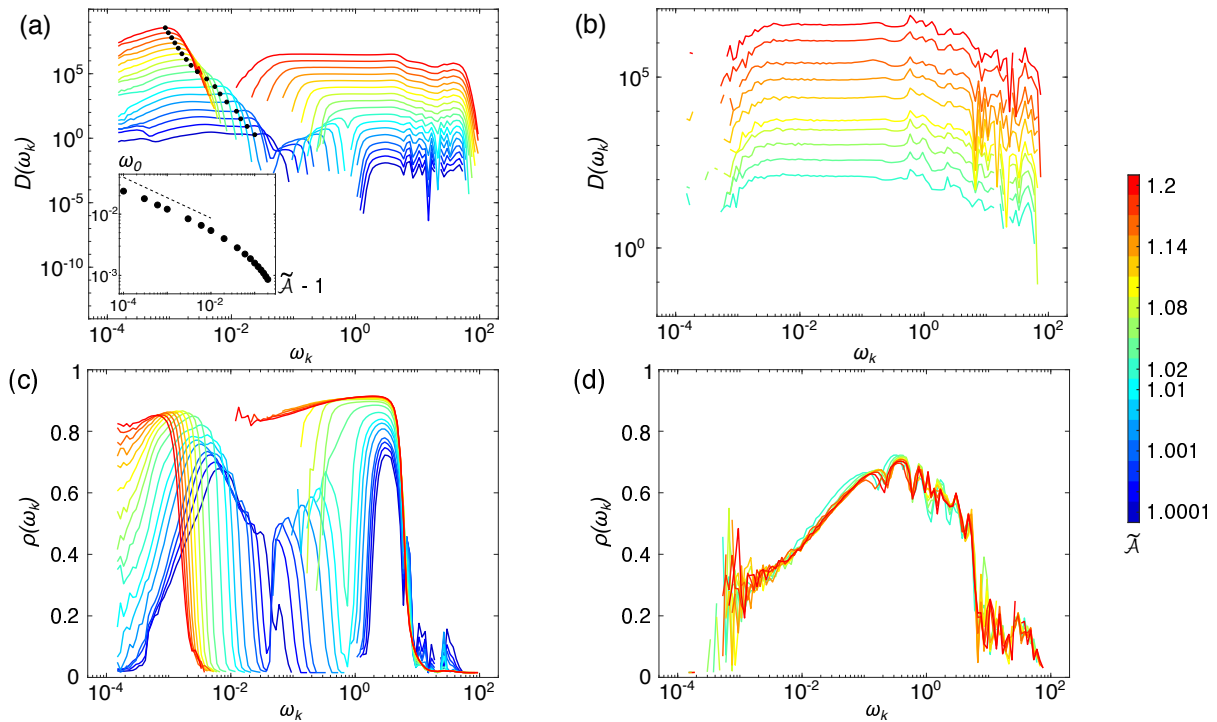}
\caption{Density of vibrational frequencies $D(\omega_k)$ for $N = 64$ jammed packings of deformable particles with (a) $\epsilon_b/\epsilon_v=0$ and (b) $10^{-3}$ over a range of  $\widetilde{\mathcal{A}}$ from $1$ (cyan) to $1.2$ (magenta). Curves in (a) and (b) are shifted up by $0.5$ in the log scale between consecutive $\mathcal{A}$ values. Black dots in (a) indicate the average quartic eigenmode frequency $\omega_0$, which is also shown in the inset to (a) as a function of $\widetilde{\mathcal{A}}-1$. The dashed line has slope $-1/3$ in the inset to (a). The participation ratio $\rho(\omega_k)$ is shown for $N=64$ jammed packings of deformable particles with (c) $\epsilon_b/\epsilon_v =0$ and (d) $10^{-3}$ over the same range of ${\widetilde {\cal A}}$. In all panels, the data are averaged over $500$ packings.}
    \label{fig:dos}
\end{figure*}

In Fig.~\ref{fig:dos} (a), we display the density of vibrational frequencies $D(\omega_k)$ for jammed packings of deformable particles with $\epsilon_b=0$ over a wide range of ${\widetilde {\cal A}}$. We find several key features in $D(\omega_k)$: 1) there is a large gap that separates the quartic and quadratic frequency bands; 2) the quartic band shifts to lower frequencies with increasing ${\widetilde {\cal A}}$; and 3) the high-frequency part of the quadratic band is insensitive to ${\widetilde {\cal A}}$, while the low-frequency part forms a plateau that extends to lower frequencies with increasing ${\widetilde {\cal A}}$. In the inset of Fig.~\ref{fig:dos} (a), we plot the average quartic mode frequency $\omega_0$ as a function of $\widetilde{\mathcal{A}}$. We find that $\omega_0 \sim (\mathcal{A}-1)^{-1/3}$ displays power-law scaling with a scaling exponent, $-1/3$, that is similar to that observed for 2D packings of deformable particles with $\epsilon_b = 0$~\cite{treado21prm}.  However, the scaling exponent is different (even the opposite sign) from the value ($1/2$) that has been observed for quartic modes in jammed packings of rigid non-spherical particles\cite{schreck12pre, brito18pnas}. 

We display $D(\omega_k)$ for jammed packings of deformable particles with non-zero $\epsilon_b$ over a range of ${\widetilde {\cal A}}$ in Fig.~\ref{fig:dos} (b). The ${\widetilde {\cal A}}$-dependence is weak. In addition, $D(\omega_k)$ for packings of deformable particles with non-zero $\epsilon_b$ is continuous with no large frequency gaps, as has been found for jammed packings of rigid, frictionless non-spherical particles\cite{zeravcic09epl, schreck12pre}.  The lack of a frequency band gap in $D(\omega_k)$ is likely caused by the coupling of the translational, rotational, and shape degrees of freedom generated by the effective friction of the spherical vertices on each particle.

\begin{figure}
    \centering
    \includegraphics[width=\linewidth]{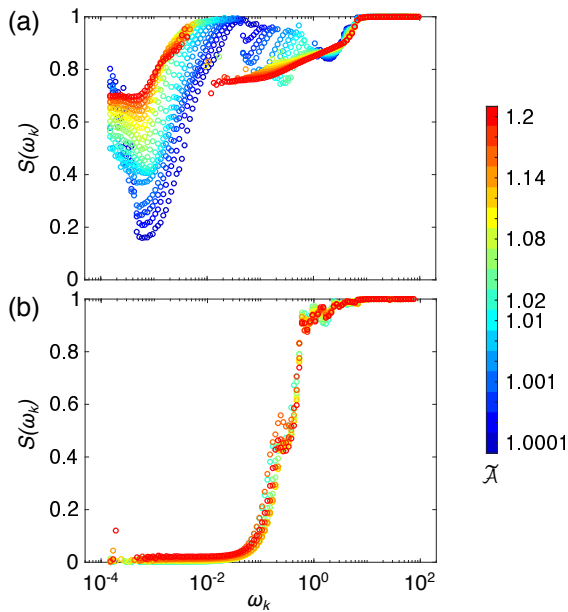}
\caption{Contribution $S(\omega_k)$ of the shape degrees of freedom to the $k$th eigenmode of the dynamical matrix (with frequency $\omega_k$) for $N=64$ jammed packings of deformable particles with (a) $\epsilon_b/\epsilon_v=0$ and (b) $10^{-3}$ over a range of  $\widetilde{\mathcal{A}}$ from $1$ (cyan) to $1.2$ (magenta).}
    \label{fig:shape}
\end{figure}

We next examine the contribution of the motion of each particle to each eigenmode of the dynamical matrix at frequency $\omega_k$ by calculating the participation ratio\cite{silbert09pre}:
\begin{equation}
    \label{eq:pr}
    \rho(\omega_k) = \frac{\left|\sum\limits_{n=1}^{N} \mathbf{e}_{\omega_k n} \cdot \mathbf{e}_{\omega_k n}\right|^2}{N\sum\limits_{n=1}^{N} \left|\mathbf{e}_{\omega_k n} \cdot \mathbf{e}_{\omega_k n}\right|^2},
\end{equation}
where $\vec{V}_k = \{\mathbf{e}_{\omega_k 1},\ldots,\mathbf{e}_{\omega_k N}\}$ and $\mathbf{e}_{\omega_k n}$ is the contribution to $\vec{V}_k$ from the $n$th particle. Small values of $\rho(\omega_k)$ indicate localized eigenmodes, whereas large values indicate spatially-extended eigenmodes. For jammed packings of deformable particles with $\epsilon_b =0$, $\rho(\omega_k)$ is complex; for a single value of ${\widetilde {\cal A}}$, it increases and decreases multiple times as the frequency increases and it depends strongly on ${\widetilde {\cal A}}$. (See Fig.~\ref{fig:dos} (c).) Interestingly, for quartic modes, $\rho(\omega_k)$ at the lowest frequency increases from $\sim 0$ to $\sim 0.8$ as $\widetilde{\mathcal{A}}$ increases from $1$ to $1.2$. This result suggests that the lowest frequency quartic modes become increasingly de-localized as jammed packings of completely deformable particles approach confluence. In contrast, for jammed packings of deformable particles with non-zero bending energy, $\rho(\omega_k)$ does not depend on $\widetilde{\mathcal{A}}$ as shown in Fig.~\ref{fig:dos} (d). In this case, $\rho(\omega_k)$ is small at both small and large $\omega_k$, suggesting localized eigenmodes occur at these frequencies, and $\rho(\omega_k)$ reaches a peak value of $\sim 0.7$ at $\omega_k \sim 10^{-1}$. This behavior for $\rho(\omega_k)$ is similar to that found for jammed packings of frictionless disks and spheres\cite{silbert09pre, xu10epl}, even though the degrees of freedom are different in these two cases.

\subsection{Contribution of shape degrees of freedom to vibrational modes}
\label{sec:shape}

To understand the role of particle deformability in the vibrational response, we decompose each eigenmode $k$ (with frequency $\omega_k$) of the dynamical matrix into contributions from the translational $T(\omega_k)$, rotational $R(\omega_k)$, and shape $S(\omega_k)$ degrees of freedom, such that $T(\omega_k) +R(\omega_k) + S(\omega_k)=1$. Details about how to calculate the eigenmode projections can be found in Appendix~A. Each projection $T(\omega_k)$, $R(\omega_k)$, and $S(\omega_k)$ varies from $0$ to $1$, with $0$ indicating no contribution of the translational, rotational, or shape degrees of freedom to the eigenmode and $1$ indicating that only translation, rotation, or shape change contributes to a given eigenmode.

\begin{figure}
    \centering
    \includegraphics[width=\linewidth]{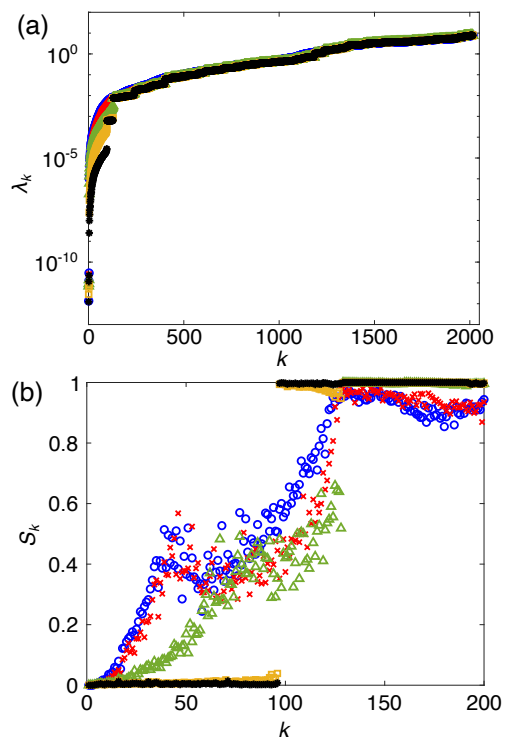}
\caption{(a) Eigenvalues of the dynamical matrix $\lambda_k$, sorted from smallest to largest, for $N = 16$ jammed packings of deformable particles with $\epsilon_b/\epsilon_v =10^{-3}$, $\widetilde{\mathcal{A}} = 1.06$, and several values of $\epsilon_c/\epsilon_v$: $1$ (blue circles), $10^{-1}$ (red crosses), $10^{-2}$ (green triangles), $10^{-3}$ (yellow squares), and $10^{-4}$ (black asterisks). (b) Magnitude of the projection of the shape degrees of freedom onto the $k$th eigenmode of the dynamical matrix $S_k$ for the same data in (a).}
    \label{fig:rigidlimit}
\end{figure}

In Fig.~\ref{fig:shape} (a), we show that for jammed packings of deformable particles with $\epsilon_b = 0$ the shape contribution $S(\omega_k)$ is non-zero over the full range of $\omega_k$ for all shape parameters $1 < \widetilde{\mathcal{A}} < 1.2$.  This result suggests that particle shape deformability plays an important role in the vibrational response for jammed packings of deformable particles. We also find that $S(\omega_k)$ increases with $\widetilde{\mathcal{A}}$ for the lowest frequencies.  The jammed packings become ``confluent" for ${\widetilde {\cal A}} \gtrsim 1.16$, and in this regime particle translations and rotations cost more energy than shape changes at low frequencies. In Fig.~\ref{fig:shape} (a), we also show that $S(\omega_k) \gtrsim 0.6$ at intermediate frequencies above the quartic mode frequencies. This result clearly distinguishes these intermediate frequency modes from those in jammed packings of frictionless, rigid non-spherical particles mainly associated with rotational degrees of freedom~\cite{schreck12pre}. In contrast, for jammed packings of deformable particles with nonzero $\epsilon_b$, $S(\omega_k) \sim 0$ at low $\omega_k$, as shown in Fig.~\ref{fig:shape} (b). $S(\omega_k)$ only becomes appreciable for $\omega_k \gtrsim 10^{-1}$. In addition, $S(\omega_k)$ does not vary significantly with $\widetilde{\mathcal{A}}$ for packings of deformable particles with nonzero $\epsilon_b$.

\begin{figure*}
    \centering
    \includegraphics[width=\textwidth]{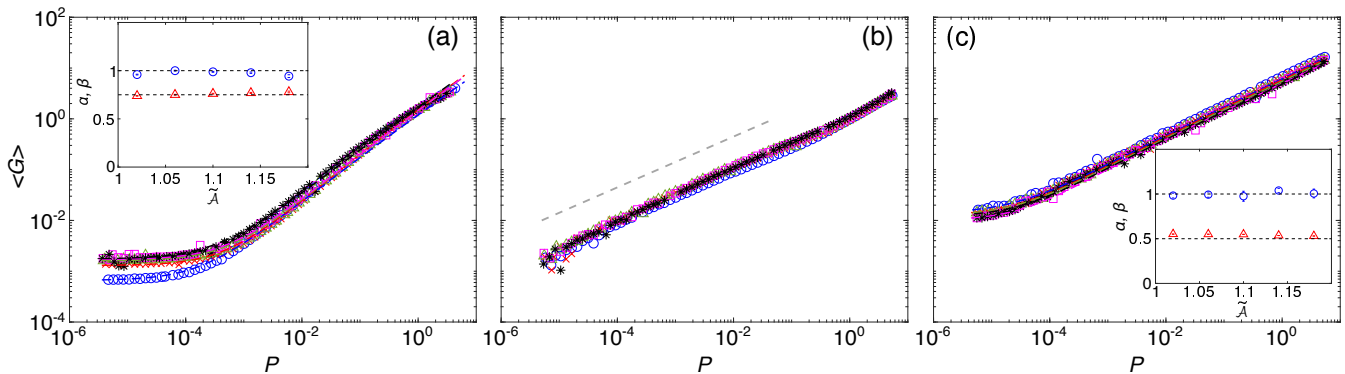}
\caption{Ensemble-averaged shear modulus $\langle G\rangle$ plotted versus pressure $P$ for $N = 128$ jammed packings of deformable particles with (a) $\epsilon_b/\epsilon_v=0$ and (b) $10^{-3}$, and of (c) rigid bumpy particles, for several shape parameters: $\widetilde{\mathcal{A}} = 1.02$ (blue circles), $1.06$ (red crosses), $1.1$ (green triangles), $1.14$ (magenta squares), and $1.18$ (black asterisks). The dashed lines in (a) and (c) indicate fits to Eq.~\ref{eq:pg} and the power-law scaling exponents, $\alpha$ (blue circles) and $\beta$ (red triangles), are shown in the insets. The dashed line in (b) has a slope of $0.5$. The data is obtained by averaging over $500$ configurations.}
    \label{fig:shearmodulus}
\end{figure*}

We now investigate how to take the rigid-particle limit for jammed packings of deformable particles to recover eigenmodes that only contain contributions from particle rotations and translations, not particle shape changes. To address this question, we study jammed packings of deformable particles with non-zero bending energy as a function of decreasing $\epsilon_c/\epsilon_v$ and fixed $\epsilon_b/\epsilon_v=10^{-3}$. In Fig.~\ref{fig:rigidlimit} (a), we show the eigenvalue spectrum of the dynamical matrix sorted from smallest to largest for $N=16$ jammed packings with ${\widetilde {\cal A}}=1.06$. As $\epsilon_c/\epsilon_v$ decreases, a band gap emerges that separates $6N - 3$ small eigenvalues from the larger band of high-frequency eigenvalues. (Note that the smallest three eigenvalues correspond to rigid translations of the system.) In Fig.~\ref{fig:rigidlimit} (b), we show that the contribution to the eigenmodes from the shape degrees of freedom, $S_k = 0$, for the first $6N$ eigenmodes for  $\epsilon_c/\epsilon_v \gtrsim 10^{-3}$. (We index the eigenmodes by the integer $k$ instead of $\omega_k$, so that it is easy to identify the first $6N$ eigenmodes.) Thus, the first $6N$ eigenmodes are composed of only particle translations and rotations, similar to the eigenmodes of jammed packings of rigid bumpy particles.

\subsection{Ensemble-averaged shear modulus}
\label{sec:shearmodulus}

In this section, we examine the effects of particle deformability on the mechanical properties of jammed packings of deformable particles. In particular, we isotropically compress the packings and calculate the ensemble-averged shear modulus $\langle G\rangle$ as a function of pressure $P$ for particles with $\epsilon_b/\epsilon_v =0$ and $10^{-3}$ and rigid bumpy particles. We find that $\langle G(P) \rangle$ can be fit by the following functional form:
\begin{equation}
    \label{eq:pg}
    \langle G(P) \rangle = G_0 + \frac{aP^\alpha}{1 + cP^{\alpha - \beta}},
\end{equation}
where $a$ and $c$ are constants, and $\alpha$ and $\beta$ are the power-law scaling exponents at small and large pressures, respectively\cite{vanderwerf20prl}. $G_0\sim N^{-1}$ gives the value of the shear modulus in the zero-pressure limit. Prior studies of jammed packings of frictionless and frictional disks in 2D and spheres in 3D have found that $\alpha \approx 1$ and $\beta \approx 0.5$\cite{ohern03pre,softp:GoodrichPRL2012,somfai07pre}. In Fig.~\ref{fig:shearmodulus} (a), we show that $\langle G\rangle$ obeys Eq.~\ref{eq:pg} for all ${\widetilde {\cal A}}$ for jammed packings of deformable particles with $\epsilon_b =0$. We find that $G_0$ decreases as ${\widetilde {\cal A}}$ approaches unity (because of the decrease in $z_J$ as ${\widetilde {\cal A}} \rightarrow 1$), but the power-law scaling exponents $\alpha \approx 1$ and $\beta \approx 0.75$ (shown in the insets) are insensitive to ${\widetilde {\cal A}}$.  Note that the power-law scaling exponent $\beta$ is different for jammed packings of completely deformable particles compared to that for rigid spherical frictionless and frictional particles ($\beta \approx 0.5$), as well as rigid, frictionless ellipse- ($1.0$)\cite{schreck10softmatter} and circulo-line-shaped particles ($0.8$-$0.9$)\cite{jerry}.  The larger values of $\beta > 0.5$ is correlated with the presence of quartic eigenmodes of the dynamical matrix. 

For jammed packings of deformable particles with nonzero $\epsilon_b$ (and no quartic eigenmodes), we do not observe a low-pressure plateau in $\langle G\rangle$ (due to the relatively small value of $\epsilon_c/\epsilon_v$), and $\langle G\rangle \sim P^{0.5}$ over the full range of pressure studied and for all $\widetilde{\mathcal{A}}$. (See Fig.~\ref{fig:shearmodulus} (b).) As a comparison,  we show $\langle G(P)\rangle$ for jammed packings of rigid bumpy particles over the same range of ${\widetilde {\cal A}}$ in Fig.~\ref{fig:shearmodulus} (c). Similar to jammed packings of deformable particles with non-zero bending energy, the power-law scaling exponent $\beta \approx 0.5$ and $\langle G(P)\rangle$ is insensitive to $\widetilde{\mathcal{A}}$.

\section{Conclusions and future directions}
\label{sec:conclusions}

In this article, we performed computational studies of the structural, vibrational, and mechanical properties of jammed packings of deformable particles in three dimensions (3D).  We first considered the vibrational response of single deformable particles with no bending energy and showed that they possess numerous unconstrained degrees of freedom.  Adding a bending energy term for each edge between triangular faces on the polyhedral surface of the particle constrains all of the remaining degrees of freedom. We then show that jammed packings of completely deformable particles with zero bending energy are hypostatic and possess $N_q$ quartic eigenmodes of the dynamical matrix, where $N_q$ matches the number of missing contacts relative to the isostatic value.  In contrast, jammed packings of deformable particles with non-zero bending energy are isostatic with no quartic eigenmodes. This result in 3D is significantly different than that in 2D. Jammed packings of deformable particles with non-zero bending energy in 2D can be hypstatic or isostatic depending on whether the particles are buckled or not. 

The density of vibrational modes $D(\omega_k)$ for packings of completely deformable particles in 3D possesses a frequency band gap between the quartic and higher frequency modes. The average quartic eigenmode frequency scales as $\omega_0 \sim (\widetilde{\mathcal{A}} - 1)^{-1/3}$, which is different than the scaling behavior of the quartic modes in jammed packings of rigid, frictionless non-spherical particles ($\omega_0 \sim ({\widetilde {\cal A}}-1)^{1/2}$).  $D(\omega_k)$ does not depend on the shape parameter ${\widetilde {\cal A}}$ for jammed packings of deformable particles with nonzero bending energy. In this case, $D(\omega_k)$ is similar to that for jammed packings of rigid, spherical particles with a plateau that extends to lower frequencies with decreasing pressure.  We also investigate the effect of particle deformability on the mechanical properties of jammed packings of deformable particles. Specifically, we calculate the ensemble-averaged shear modulus $\langle G\rangle$ as a function of pressure $P$ as we isotropically compress the system above jamming onset. We find that for particles with non-zero bending energy $\langle G(P)\rangle$ scales as a power-law in pressure, $\langle G\rangle \sim P^{\beta}$ with $\beta \sim 0.5$, which is similar to the results for jammed packings of rigid, frictionless and frictional spherical particles. 
The scaling behavior of the ensemble-averaged shear modulus is different for jammed packings of completely deformable particles with $\epsilon_b =0$.  In this case, the power-law scaling exponent $\beta \approx 0.75$.  Moreover, in all cases studied for which jammed particle packings (with repulsive linear spring interactions) possess quartic eigenmodes of the dynamical matrix, the power-law scaling exponent $\beta > 0.5$. For example,  
$\beta \approx 0.75$ for 2D and 3D jammed packings of completely deformable particles\cite{treado21prm}, $\beta \approx 1.0$ for 2D jammed packings of ellipse-shaped particles, and $\beta \approx 0.8$-$0.9$ for 2D jammed packings of circulo-lines\cite{jerry}.  We encourage future studies to understand the link between quartic eigenmodes of the dynamical matrix and the non-trivial power-law scaling of $\langle G(P)\rangle$.

In summary, we have shown that particle shape deformability has a significant impact on the structural, vibrational, and mechanical properties of jammed particle packings.  In the current studies, we used spherical vertices on the particle surfaces, i.e. the rough surface model, to implement the particle-particle interactions. In future studies, we will investigate the smooth surface model, where deformable particles are modelled as sphero-polyhedra, and inter-particle distances are determined by the separations between points, lines, and planes that form the particle surfaces.  It will be interesting to determine whether any of the properties of jammed packings of deformable particles depend on the surface roughness.  In addition, the current studies have determined the properties of jammed packings of deformable particles at {\it zero temperature}.  An important topic of future research is to understand how the structural, vibrational, and mechanical properties depend on temperature, and how the glass transition temperature that determines long-time particle diffusion depends on the shape parameter and bending rigidity\cite{activesphere:Omar21prl}.

\section*{Acknowledgements}

We acknowledge support from NSF Grants No. CMMI-2029756 (J.D.T. and C.S.O.), No. CBET-2002782 (J.D.T. and C.S.O.), and No. CBET-2002797 (M.D.S.), and NIH Award No. 5U54CA210184-04 (D.W.). This work was also supported by the High Performance Computing facilities operated by Yale’s Center for Research Computing.

\section*{Appendix A}
\label{sec:modeproj}

In this Appendix, we describe how to decompose the eigenmodes of the dynamical matrix into contributions from the translational, rotational, and shape degrees of freedom. We consider a packing of $N$ deformable particles, where each particle $n$'s center of mass is located at $\vec{c}_n = N_v^{-1}\sum_{i=1}^{N_v} \vec{r}_{in}$. Let $\vec{V}^k$ be the $k$th eigenvector of the dynamical matrix $M$ in Cartesian coordinates. Components from the $(3N_v(n-1)+1)$th to the $(3N_vn)$th position in $\vec{V}^j$ correspond to the $n$th deformable particle, among which the first, second, and third $N_v$ components are the $N_v$ $x$-, $y$-, and $z$-coordinates, respectively. We can define six unit vectors to describe translation ($\vu{u}_{n,x}$, $\vu{u}_{n,y}$, $\vu{u}_{n,z})$ and rotation ($\vu{u}_{n,r1}$, $\vu{u}_{n,r2}$, $\vu{u}_{n,r3}$) about the center of mass of the $n$th particle as follows: 
\begin{equation}
\begin{split}
    \vu{u}_{n, x} = \frac{\vec{u}_{n, x}}{|\vec{u}_{n, x}|}, \vec{u}_{n, x} = & (\underbrace{0, \ldots, 0}_{\text{1 to ($n - 1$)}}, \underbrace{1, \ldots, 1}_{\text{$n$th particle $x$}}, \\ & \underbrace{0, \ldots, 0}_{\text{$n$th particle $y$ and $z$}}, \underbrace{0, \ldots, 0}_{\text{($n$+1) to $N$}}),
\end{split}
\end{equation}
\begin{equation}
\begin{split}
    \vu{u}_{n, y} = \frac{\vec{u}_{n, y}}{|\vec{u}_{n, y}|}, \vec{u}_{n, y} = & (\underbrace{0, \ldots, 0}_{\text{$1$ to ($n - 1$) }}, \underbrace{0, \ldots, 0}_{\text{$n$th particle $x$}}, \underbrace{1, \ldots, 1}_{\text{$n$th particle $y$}}, \\ &\underbrace{0, \ldots, 0}_{\text{$n$-th particle $z$}}, \underbrace{0, \ldots, 0}_{\text{$(n+1)$ to $N$}}),
\end{split}
\end{equation}
\begin{equation}
\begin{split}
    \vu{u}_{n, z} = \frac{\vec{u}_{n, z}}{|\vec{u}_{n, z}|}, \vec{u}_{n, z} = & (\underbrace{0, \ldots, 0}_{\text{$1$ to ($z - 1$) }}, \underbrace{0, \ldots, 0}_{\text{$n$th particle $x$ and $y$}}, \\ & \underbrace{1, \ldots, 1}_{\text{$n$th particle $z$}}, \underbrace{0, \ldots, 0}_{\text{$(n+1)$ to $N$}}),
\end{split}
\end{equation}
\begin{equation}
\begin{split}
    \vu{u}_{n, r1} = \frac{\vec{u}_{n, r1}}{|\vec{u}_{n, r1}|}, \vec{u}_{n, r1} = & (\underbrace{0, \ldots, 0}_{\text{$1$ to ($n - 1$) }}, \underbrace{0, \ldots, 0}_{\text{$n$th particle $x$}}, \\ & \underbrace{-(z_{1n} - c_{n, z}), \ldots, -(z_{N_vn} - c_{n, z})}_{\text{$n$th particle $y$}}, \\ & \underbrace{y_{1n} - c_{n, y}, \ldots, y_{N_vn} - c_{n, y}}_{\text{$n$th particle $z$}},  \\ & \underbrace{0, \ldots, 0}_{\text{$(n+1)$ to $N$}}),
\end{split}
\end{equation}
\begin{equation}
\begin{split}
    \vec{u}'_{n, r2} = & (\underbrace{0, \ldots, 0}_{\text{$1$ to ($n - 1$) }}, \underbrace{-(z_{1n} - c_{n, z}), \ldots, -(z_{N_vn} - c_{n, z})}_{\text{$n$th particle $x$}}, \\ & \underbrace{0, \ldots, 0}_{\text{$n$th particle $y$}}, \underbrace{x_{1n} - c_{n, x}, \ldots, x_{N_vn} - c_{n, x}}_{\text{$n$th particle $z$}}, \\ &  \underbrace{0, \ldots, 0}_{\text{$(n+1)$ to $N$}}),
\end{split}
\end{equation}
and
\begin{equation}
\begin{split}
    \vec{u}'_{n, r3} = & (\underbrace{0, \ldots, 0}_{\text{$1$ to ($n - 1$) }}, \underbrace{-(y_{1n} - c_{n, y}), \ldots, -(y_{N_vn} - c_{n, y})}_{\text{$n$-th particle $x$}}, \\ & \underbrace{x_{1n} - c_{n, x}, \ldots, x_{N_vn} - c_{n, x}}_{\text{$n$-th particle $y$}}, \underbrace{0, \ldots, 0}_{\text{$n$-th particle $z$}}, \\ & \underbrace{0, \ldots, 0}_{\text{$(n+1)$ to $N$}}).
\end{split}
\end{equation}
Note that these six vectors do not form an orthogonal basis due to non-zero off-diagonal components in the moment of inertia matrix with respect to the center of mass. To construct six orthogonal unit vectors, we apply the Gram-Schmidt process ($\vu{u}_{n, x}$, $\vu{u}_{n, y}$, $\vu{u}_{n, z}$, and $\vu{u}_{n, r1}$ are already orthogonal to each other):
\begin{equation}
\begin{split}
    \vu{u}_{n, r2} = \frac{\vec{u}_{n, r2}}{|\vec{u}_{n, r2}|}, \vec{u}_{n, r2} = & \vec{u}'_{n, r2} - \frac{\vec{u}'_{n, r2} \cdot \vu{u}_{n, x}}{|\vec{u}'_{n, r2}|} \vu{u}_{n, x} \\ & - \frac{\vec{u}'_{n, r2} \cdot \vu{u}_{n, y}}{|\vec{u}'_{n, r2}|} \vu{u}_{n, y} \\ & - \frac{\vec{u}'_{n, r2} \cdot \vu{u}_{n, z}}{|\vec{u}'_{n, r2}|} \vu{u}_{n, z} \\ & - \frac{\vec{u}'_{n, r2} \cdot \vu{u}_{n, r1}}{|\vec{u}'_{n, r2}|} \vu{u}_{n, r1}
\end{split}
\end{equation}
and
\begin{equation}
\begin{split}
    \vu{u}_{n, r3} = \frac{\vec{u}_{n, r3}}{|\vec{u}_{n, r3}|}, \vec{u}_{n, r3} = & \vec{u}'_{n, r3} - \frac{\vec{u}'_{n, r3} \cdot \vu{u}_{n, x}}{|\vec{u}'_{n, r3}|} \vu{u}_{n, x} \\ & - \frac{\vec{u}'_{n, r3} \cdot \vu{u}_{n, y}}{|\vec{u}'_{n, r3}|} \vu{u}_{n, y} \\ & - \frac{\vec{u}'_{n, r3} \cdot \vu{u}_{n, z}}{|\vec{u}'_{n, r3}|} \vu{u}_{n, z} \\ & - \frac{\vec{u}'_{n, r3} \cdot \vu{u}_{n, r1}}{|\vec{u}'_{n, r3}|} \vu{u}_{n, r1} \\ & - \frac{\vec{u}'_{n, r3} \cdot \vu{u}_{n, r2}}{|\vec{u}'_{n, r3}|} \vu{u}_{n, r2}.
\end{split}
\end{equation}
By defining the following coefficients,
\begin{align}
    p_{n, x}^k &= \vec{V}^k \cdot \vu{u}_{n, x}\\
    p_{n, y}^k &= \vec{V}^k \cdot \vu{u}_{n, y}\\
    p_{n, z}^k &= \vec{V}^k \cdot \vu{u}_{n, z}\\
    p_{n, r1}^k &= \vec{V}^k \cdot \vu{u}_{n, r1}\\
    p_{n, r2}^k &= \vec{V}^k \cdot \vu{u}_{n, r2}\\
    p_{n, r3}^k &= \vec{V}^k \cdot \vu{u}_{n, r3},
\end{align}
we can rewrite the eigenvector $\vec{V}^k$ as
\begin{equation}
\begin{split}
    \vec{V}^k = & \sum_{n = 1}^{N} p_{n, x}^k \vu{u}_{n, x} + \sum_{n = 1}^{N} p_{n, y}^k \vu{u}_{n, y} + \sum_{n = 1}^{N} p_{n, z}^k \vu{u}_{n, z} \\ & + \sum_{n = 1}^{N} p_{n, r1}^k \vu{u}_{n, r1} + \sum_{n = 1}^{N} p_{n, r2}^k \vu{u}_{n, r2} \\ & + \sum_{n = 1}^{N} p_{n, r3}^k \vu{u}_{n, r3} + \vec{V}_s^k,
\end{split}
\end{equation}
where $\vec{V}_s^k$ is the vector that remains after projecting the particle translations and rotations out of $\vec{V}^k$. By applying this decomposition, we can express each eigenmode as the sum of particle translations, rotations, and shape deformations.

\begin{figure}
    \includegraphics[width=\linewidth]{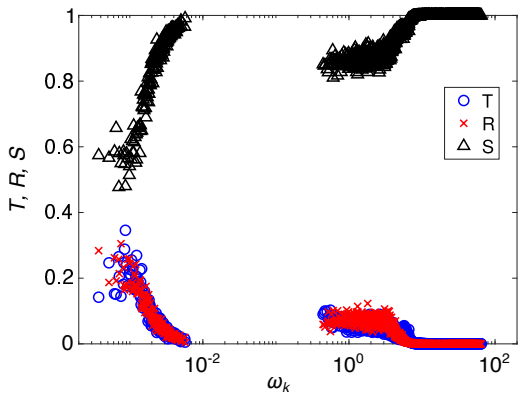}
\caption{Magnitude of the contributions of the translational $T$ (blue circles), rotational $R$ (red crosses), and shape $S$ (black triangles) degrees of freedom to each eigenmode of the dynamical matrix with frequency $\omega_k$ for $N = 16$ jammed packings of deformable particles with $\epsilon_b=0$ and $\widetilde{\mathcal{A}} = 1.06$.}
    \label{fig:decomp}
\end{figure}

With these coefficients, we can define the contributions of translational $T^k$ and rotational $R^k$ degrees of freedom to the $k$th eigenmode of the dynamical matrix as:
\begin{align}
    T^k &= \sum_{n = 1}^{N} \qty[\qty(p_{n, x}^k)^2 + \qty(p_{n, y}^k)^2 + \qty(p_{n, z}^k)^2]\\
    R^k &= \sum_{n = 1}^{N} \qty[\qty(p_{n, r1}^k)^2 + \qty(p_{n, r2}^k)^2 + \qty(p_{n, r3}^k)^2].
\end{align}
$S^k = 1 - T^k - R^k$ gives the contribution of the shape degrees of freedom to the $k$th eigenmode. As an example, we show $T^k, R^k$, and $S^k$ for an $N=16$ jammed packing of deformable particles with $\epsilon_b/\epsilon_v = 0$ with shape parameter $\widetilde{\mathcal{A}} = 1.06$ in Fig.~\ref{fig:decomp} as a function of frequency $\omega_k$.

\section*{Appendix B}
\label{sec:pointstessellation}

In this Appendix, we provide insight into the value of the shape parameter at which jammed packings of deformable particles with zero bending energy become confluent. 
In particular, we show results for the probability distribution of the shape parameters obtained from Voronoi tessellation of random points and of the sphere centers in jammed packings of monodisperse, frictionless spheres, as well as the shape parameters of the polyhedra generated from Lloyd's algorithm\cite{lloyd}.  In all three cases, we consider $N_p=64$ points and periodic boundary conditions in the $x$-, $y$-, and $z$-directions. For the jammed sphere packings, we use the same packing-generation process described in Sec.~\ref{sec:methods}. For Lloyd's algorithm, we start with a set of random points and apply Voronoi tessellation. We then use the centroids of the tessellated polyhedra as the new set of points and apply Voronoi tessellation again. We repeat this process $10^4$ times after which the distribution of the polyhedra shape parameters, ${\cal P}({\cal A})$, reaches a stationary distribution. 

\begin{figure}
    \includegraphics[width=\linewidth]{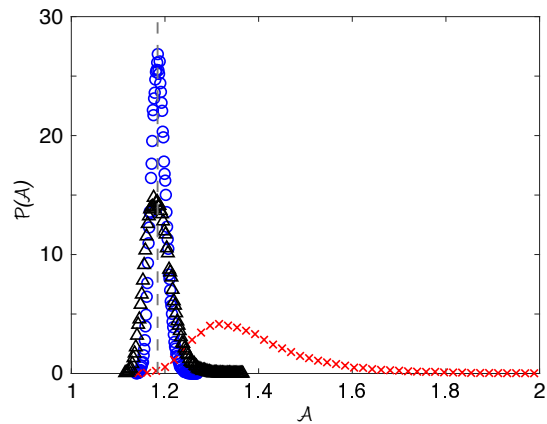}
\caption{The probability distribution $\mathcal{P}(\mathcal{A})$ of the polyhdedra generated from Voronoi-tessellating the centers of spheres in jammed monodisperse sphere packings (blue circles) and random points (red crosses), as well as the polyhedra generated from Lloyd's algorithm (black triangles). In all cases, the number of points is $N_p=64$ with periodic boundary conditions in the $x$-, $y$-, and $z$-directions. The vertical dashed line is located at $\mathcal{A} = 1.18$.}
    \label{fig:pointsp}
\end{figure}

In Fig.~\ref{fig:pointsp}, we show ${\cal P}({\cal A})$ 
for the three point processes described above. The distributions ${\cal P}({\cal A})$ from jammed frictionless sphere packings and Lloyd's algorithm are narrow with peaks near $\mathcal{A} \approx 1.185 = 1.157\mathcal{A}_v$ and $\approx 1.176 = 1.148\mathcal{A}_v$, respectively. As discussed in Sec.~\ref{sec:methods}, $\mathcal{A}_v = 1.024$ is the smallest shape parameter for the $N_v = 42$ polyhedral deformable particles that we consider in the main text. Thus, the most probable shape parameters for these two types of Voronoi tessellations are similar to the value of $\widetilde{\mathcal{A}}={\cal A}/{\cal A}_v \approx 1.16$ above which the packing fraction at jamming onset $\phi_J$ reaches a plateau for deformable particles with $\epsilon_b = 0$. This value of ${\cal A}$ is also similar to the critical shape parameter at which a fliud-to-solid transition occurs in the 3D vertex model for confluent tissues~\cite{merkel18iop}. In contrast, Voronoi tessellations obtained from sets of random points yield a wide distribution ${\cal P}({\cal A})$ with the most probable $\mathcal{A} \approx 1.316 = 1.285\mathcal{A}_v$, which is much larger than the most probable values from the other two types of Voronoi tessellations.

\section*{Appendix C}

\begin{figure}
    \centering
    \includegraphics[width=\linewidth]{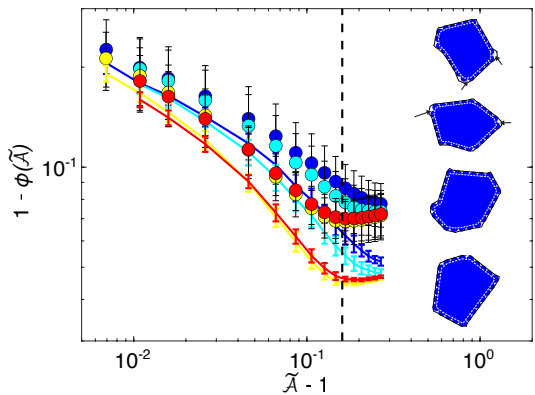}
    \caption{Local (symbols) and global (lines) packing fraction $1-\phi$ versus particle shape parameter ${\widetilde {\cal A}}-1$ for jammed packings of deformable particles in 2D prepared using a packing-generation protocol that includes thermal fluctuations at temperatures $T = 10^{-6}$ (blue), $10^{-4}$ (cyan), $10^{-3}$ (yellow), and $10^{-2}$ (red). Error bars correspond to averages over configurations (lines) or both particles and configurations (symbols). The vertical dashed line is drawn at $\widetilde{\mathcal{A}} = 1.16$. Example particles with $\widetilde{\mathcal{A}} = 1.16$ for their surface-Voronoi cells are drawn in the inset, sorted by increasing temperature from top to bottom. Arrows indicate excess free area in each cell. Surface-Voronoi cells are drawn with black solid lines, while total particle areas are shaded in blue, and the underlying polygons of the deformable particles are indicated with white dashed lines. }
    \label{fig:voro2D}
\end{figure}

In this Appendix, we describe the effects on the packing fraction at jamming onset $\phi_J$ in 2D from packing-generation protocols that include thermal fluctuations. In previous studies~\citep{boromand18prl}, we found that $\phi_J$ increases with shape parameter for ${\widetilde {\cal A}} < {\widetilde {\cal A}}^{\dagger} \approx 1.16$. Above ${\widetilde {\cal A}}^{\dagger}$, $\phi_J$ reaches a plateau and the particle shapes begin to buckle inward. As the polygons of Voronoi tessellations of jammed disk packings possess typical shape parameter of $\mathcal{A}^{\dagger} \approx 1.16$, we hypothesized that the plateau in $\phi_J$ for ${\widetilde {\cal A}} \gtrsim {\widetilde {\cal A}}^{\dagger}$ indicates a confluence transition, where deformable particles completely fill their Voronoi cells as ${\widetilde{\cal A}} \to {\widetilde {\cal A}}^{\dagger}$. For ${\widetilde {\cal A}} > {\widetilde {\cal A}}^{\dagger}$, the particles cannot further expand in area to increase their perimeter, so they invaginate instead.

We show in Fig.~\ref{fig:voro2D} that the confluence transition is sensitive to the packing-generation protocol. We prepare jammed packings of $N = 64$ 2D deformable particles with $\epsilon_b=0$ in square, periodic boundaries with side length $L$. To include thermal fluctuations in the packing-geneartion proptocol, we run constant $N$, constant boundary area $L^2$, and constant temperature $T$ dynamics for a time $50 \tau$, where $\tau = \sqrt{a_0/\epsilon_c}$, $a_0$ is the preferred area of the particle, and thermal energy $k_B T$ is given in units of $\epsilon_c$. We then rapidly quench the system to $T = 0$ using FIRE, take a small compression step, and then re-minimize the total potential energy to achieve force balance. We repeat this thermalization, compression, and energy minimization process until reaching jamming onset with a pressure that satisfies $10^{-7} < P < 2 \times 10^{-7}$ when the system is in force balance. (A similar protocol has been implemented to generate jammed packings of 3D rigid bumpy particles~\cite{mei20proteins}.)   We studied a range of temperatures from $T=10^{-6}$ to $10^{-2}$. Constant temperature was enforced using a Langevin thermostat~\citep{sim:AllenOxford2017}.

We measure packing fraction both globally and locally; the global packing fraction $\phi = L^{-2}\sum_\mu a_{t\mu}$, where $a_{t\mu}$ is the \emph{total} area of particle $\mu$, i.e. the area of the underlying polygon $a_\mu$ plus the area of the exposed bumpy vertices $a_{b\mu}$. For a particle with $n_\mu$ circular vertices of radius $r_\mu$, the exposed bump area $a_{b\mu} = \qty(\frac{n_\mu}{2} - 1)\pi r_\mu^2$. The local packing fraction for particle $\mu$ is defined as $\phi_\mu = a_{t\mu} / a_{v\mu}$, where $a_{v\mu}$ is the area of the surrounding surface-Voronoi cell of the 2D deformable particle~\citep{voro:SchallerPhiMag2013}. Surface-Voronoi diagrams are generated by distributing fifteen points along the segments joining adjacent circular vertices on each particle, computing the Voronoi tessellation of all of the points, and taking the union of the Voronoi cells associated with each deformable particle. 

\begin{figure}
    \includegraphics[width=\linewidth]{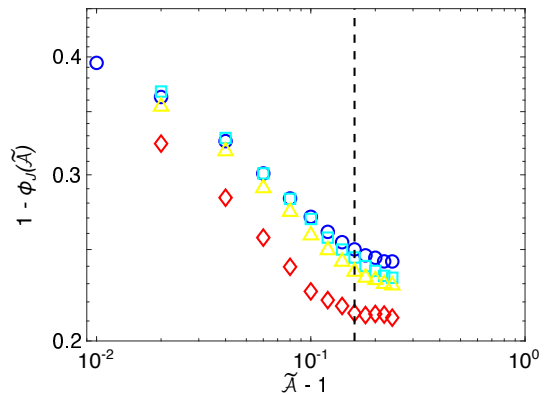}
    \caption{Packing fraction at jamming onset $1 - \phi_J$ plotted as a function of shape parameter $\widetilde{\mathcal{A}}-1$ for jammed packings of $N = 16$ deformable particles with $\epsilon_b = 0$ in $3D$, generated using the protocol with thermal fluctuations at temperature $T = 0$ (blue circles), $10^{-4}$ (cyan squares), $10^{-3}$ (yellow triangles), and $10^{-2}$ (red diamonds). The dashed vertical line is located at $\widetilde{\mathcal{A}} = 1.16$. Note that the confluence transition in 3D sharpens for packings with increasing $T$, but it is still smoother than that for packings of deformable particles in 2D as shown in Fig.~\ref{fig:voro2D}.}
    \label{fig:phi_qa}
\end{figure}

In Fig.~\ref{fig:voro2D}, we find that there is a well-defined confluence transition in both the global and local packing fractions for 2D deformable particle packings generated with large thermal fluctuations. When $T \ge 10^{-3}$, both measures of the packing fraction possess maxima near $\widetilde{\mathcal{A}} \approx 1.16$. When $T < 10^{-3}$, the packing fraction at jamming onset is generally smaller and continues to change for ${\widetilde {\cal A}} > {\widetilde {\cal A}}^{\dagger}$. In the inset of Fig.~\ref{fig:voro2D}, we include examples of single deformable particles with $\widetilde{\mathcal{A}} = 1.16$ and their associated surface-Voronoi cells. At lower temperatures, we find small regions of excess free area near the cell boundaries, but at higher temperatures these regions disappear. This result indicates that lower local (and therefore, global) packing at lower temperatures is caused by surface friction from the circular vertices.

\section*{Appendix D}
\label{sec:qapacking}

In this Appendix, we show that the packing fraction at jamming onset $\phi_J$ also depends on the protocol used to generate jammed packings of $3D$ deformable particles with $\epsilon_b = 0$. In addition to the packing-generation protocol described in Sec.~\ref{sec:methods}, we employ the  protocol~\cite{mei20proteins} with thermal fluctuations described in Appendix C.  As in 2D, we find that $\phi_J$ increases with $T$, as shown in Fig.~\ref{fig:phi_qa}. For small $T$ ($T \leq 10^{-3}$), $\phi_J$ smoothly approaches a maximum value of packing fraction that occurs for ${\widetilde {\cal A}} > {\widetilde {\cal A}}^{\dagger} \approx 1.16$. For $T = 10^{-2}$, we find a sharper transition near $\widetilde{\mathcal{A}} = \widetilde{\mathcal{A}}^{\dagger}$ for the maximum $\phi_J$, likely from reducing the surface friction between particles via thermal fluctuations. However, the confluence transition in 3D still appears to be less sharp than that in 2D.

\pagebreak

\bibliography{rsc} 

\end{document}